\documentclass{pasj01}
\draft 
\Received{$\langle$reception date$\rangle$}
\Accepted{$\langle$acception date$\rangle$}
\Published{$\langle$publication date$\rangle$}

\newcommand{\average}[1]{\ensuremath{\langle#1\rangle} }

\begin{document}

\title{A centrally concentrated sub-solar mass starless core in the Taurus L1495 filamentary complex}
\author{Kazuki \textsc{Tokuda},\altaffilmark{1,2,}$^{*}$ 
Kengo \textsc{Tachihara},\altaffilmark{3}
Kazuya \textsc{Saigo},\altaffilmark{2}
Phillipe \textsc{Andr\'e},\altaffilmark{4} 
Yosuke \textsc{Miyamoto},\altaffilmark{3} 
Sarolta \textsc{Zahorecz},\altaffilmark{1,2} 
Shu-ichiro \textsc{Inutsuka},\altaffilmark{3} 
Tomoaki \textsc{Matsumoto},\altaffilmark{5} 
Tatsuyuki \textsc{Takashima},\altaffilmark{1}
Masahiro N. \textsc{Machida},\altaffilmark{6} 
Kengo \textsc{Tomida},\altaffilmark{7} 
Kotomi \textsc{Taniguchi},\altaffilmark{8} 
Yasuo \textsc{Fukui},\altaffilmark{3} 
Akiko \textsc{Kawamura},\altaffilmark{2}
Ken'ichi \textsc{Tatematsu},\altaffilmark{9}
Ryo \textsc{Kandori},\altaffilmark{10}
and Toshikazu \textsc{Onishi}\altaffilmark{1}}%

\altaffiltext{1}{Department of Physical Science, Graduate School of Science, Osaka Prefecture University, 1-1 Gakuen-cho, Naka-ku, Sakai, Osaka 599-8531, Japan}
\altaffiltext{2}{National Astronomical Observatory of Japan, National Institutes of Natural Science, 2-21-1 Osawa, Mitaka, Tokyo 181-8588, Japan} 
\altaffiltext{3}{Department of Physics, Nagoya University, Chikusa-ku, Nagoya 464-8602, Japan}
\altaffiltext{4}{Laboratoire d’Astrophysique (AIM), CEA, CNRS, Universit\'e Paris-Saclay, Universit\'e Paris Diderot, Sorbonne Paris Cit\'e, 91191 Gif-sur-Yvette, France}
\altaffiltext{5}{Faculty of Sustainability Studies, Hosei University, Fujimi, Chiyoda-ku, Tokyo 102-8160, Japan} 
\altaffiltext{6}{Department of Earth and Planetary Sciences, Faculty of Sciences, Kyushu University, Nishi-ku, Fukuoka 819-0395, Japan} 
\altaffiltext{7}{Department of Earth and Space Science, Osaka University, Toyonaka, Osaka 560-0043, Japan}
\altaffiltext{8}{Departments of Astronomy and Chemistry, University of Virginia, Charlottesville, VA 22904, USA}
\altaffiltext{9}{National Astronomical Observatory of Japan, 2-21-1 Osawa, Mitaka, Tokyo 181-8588, Japan}
\altaffiltext{10}{Astrobiology Center of NINS, 2-21-1, Osawa, Mitaka, Tokyo 181-8588, Japan}

\email{tokuda@p.s.osakafu-u.ac.jp}


\maketitle

\begin{abstract}
The formation scenario of brown dwarfs is still unclear because observational studies to investigate its initial condition are quite limited. Our systematic survey of nearby low-mass star-forming regions using the Atacama Compact Array (aka the Morita array) and the IRAM 30\,m telescope in 1.2\,mm continuum has identified a centrally concentrated starless condensation with a central H$_2$ volume density of $\sim$10$^6$\,cm$^{-3}$, MC5-N, connected to a narrow (width $\sim$0.03 pc) filamentary cloud in the Taurus L1495 region.
The mass of the core is $\sim$\textcolor{black}{0.2--0.4}\,$M_{\odot}$, which is an order of magnitude smaller than typical low-mass prestellar cores. Taking into account a typical \textcolor{black}{core to} star formation efficiency for prestellar cores ($\sim$20\%--40\%) in nearby molecular clouds, 
\textcolor{black}{brown dwarf(s) or very low-mass star(s) may be going to be formed in this core.} We have found possible substructures at the high-density portion of the core, although much higher angular resolution observation is needed to clearly confirm them. The subsequent N$_2$H$^+$ and N$_2$D$^+$ observations using the Nobeyama 45 m telescope have confirmed the high-deuterium fractionation ($\sim$\textcolor{black}{30\%}). These dynamically and chemically evolved features indicate that this core is on the verge of proto-brown dwarf or very low-mass star formation and is an ideal source to investigate the initial conditions of such low-mass objects via gravitational collapse and/or fragmentation of the filamentary cloud complex.
\end{abstract}


\section{Introduction}
\subsection{Quest for low-mass prestellar cores on the verge of protostar formation \label{Intro:core}}
\ Observational studies of the earliest stage of star formation in molecular clouds are crucial in order to understand how dense cores collapse and fragment into protostars or protostellar systems. An evolutionary stage between the starless core and protostellar phases has been the missing link in understanding the initial condition of star formation. Detecting the innermost high-density parts and/or substructures within prestellar cores by high-angular resolution observations is key to understanding the nature of fragmentation and collapse on the verge of star formation.\\
\ Although the total number is still limited, interferometric observations in the dust continuum have been identifying further substructures within starless cores in nearby high-mass and low-mass cluster forming regions, such as $\rho$-Ophiuchus, Perseus, and Orion (e.g., \cite{Schnee10,Schnee12,Andre12,Nakamura12,Friesen14,Lis16,Kirk17,Ohashi18}).
For isolated low-mass star-forming clouds, such as Taurus, the dust continuum observations toward the innermost part of starless cores remain to be explored further by interferometers.  
Attempts have been bade with recent systematic surveys using the ALMA 12\,m array to detect high-density condensations just before protostar formation to investigate the nature of evolved dense cores and substructure formation. \citet{Dunham16} carried out a survey toward $\sim$60 starless cores with the mass range of from sub-solar to a few\,$M_{\odot}$ in the Chamaeleon I region \citep{Belloche11} using the ALMA 12\,m array, and they did not find any 3\,mm continuum detections among the observed starless cores. They suggest that most of the starless cores have not evolved to have the substructures with the density of $\gtrsim$10$^8$\,cm$^{-3}$ therein \textcolor{black}{or} the turbulent fragmentation scenario (e.g., \cite{Fisher04,Goodwin04}) may not be appropriate for the dense core evolution in this region. This result is also consistent with the fact that most of the starless cores have flat inner density profiles obtained with early single-dish studies in the dust continuum and near-infrared extinction (e.g., \cite{Ward94,Ward99,Motte01,Kirk05,Kandori05}). However, the angular resolution of single-dish telescopes is not always sufficient to confirm the presence or absence of substructures. Short-spacing baseline interferometer, such as the ACA (Atacama Compact Array, aka Morita Array), are essential to trace the density distributions from a single-dish scale of $\sim$10$\arcsec$ down to an interferometer scale continuously, as demonstrated by recent ALMA observations (e.g., \cite{Tokuda16,Ohashi18}). The ACA observations in dust continuum can be a useful tool to distinguish the (column) density enhancement in dense cores, which has not previously traced by single-dish observations. \\
\ It is also known that high-density gas tracers are useful for determining the evolutionary state of starless cores. 
With respect to the higher density interior of dense cores, the nitrogen bearing species such as N$_2$H$^+$ and NH$_3$ become the dominant tracers, and the abundance of deuterated counterparts is considered to be highly enhanced in a cold ($T$$\sim$10 K) and dense ($>$10$^5$\,cm$^{-3}$) environment, such as low-mass prestellar cores (e.g., \cite{Caselli02,Aikawa03,Crapsi05,Tatematsu17}). 
According to the analysis of a large number of starless cores in low-mass star-forming regions, the $N$(N$_{2}$D$^{+}$)/$N$(N$_{2}$H$^{+}$) ratio tends to increase with the density increase of the cores \citep{Crapsi05}. 
As high-deuterium enrichments are expected in high-density cores, observations of such species also provide us further information on the dense core evolution.

\subsection{Formation of brown dwarfs \label{Intro:BD}}
\ The formation process of brown dwarfs and/or very-lowmass stars is still under debate. For example, some earlier theoretical studies suggested that brown dwarfs are formed by the gravitational collapse of low-mass prestellar cores, the fragmentation of massive protostellar disks, the ejection of protostellar embryos from their natal cores, or the photo-erosion of pre-existing cores overrun by \textcolor{black}{H$\;${\sc ii} regions} (see \cite{Whitworth07} and references therein). Since these mechanisms are considered to not be mutually exclusive, detailed observational studies regarding the above-mentioned ideas are needed to obtain the comprehensive knowledge of the brown dwarf formation.\\
\  In this study, we mainly focus on the mechanism that a brown-dwarf is formed in a dense core in a similar manner of Sun-like stars (e.g., \cite{Machida09,Andre12}). Taking into account the similarity between the initial mass function (IMF) and the core mass function (CMF) (e.g., \cite{Motte98,Onishi02, Tachihara02}), progenitors of such very low-mass objects are considered to be very tiny prestellar cores with sub-stellar masses. Deep submillimeter observations (e.g., \cite{Nutter07}) found that the CMF turns over at $\sim$1 $M_{\odot}$ and the power law index of the low-mass side is also consistent with that of the IMF (\textcolor{black}{see also \cite{Konyves15,Marsh16}}). 
The fragmentation due to the Jeans instability may not produce such tiny condensations with the typical gas density of filamentary molecular clouds (10$^3$--10$^4$\,cm$^{-3}$) in low-mass star-forming regions (e.g., \cite{Onishi96}). However, some theoretical studies allow the formation of such tiny condensations by the fragmentation of higher density filaments after the radial collapse (\authorcite{Inutsuka92} \yearcite{Inutsuka92,Inutsuka97}).
As an alternative scenario, in turbulent molecular clouds, compressions by turbulent flows can also produce gravitationally unstable high-density regions locally (e.g., \cite{Padoan02}). In order to constrain the formation scenarios of brown dwarfs, it is necessary to reveal the physical properties of prestellar condensations on the verge of star formation.

\subsection{A dense core, MC5-N}
\ Our target dense core, MC5, was found by the large scale molecular gas survey in Taurus (Onishi et al. 1996; 1998; 2002). This core is embedded in the L1495 complex with entangled multiple velocity filaments (e.g., \cite{Mizuno95,Goldsmith08,Narayanan08,Hacar13}). \citet{Tafalla15} revealed its $``$chain$"$-like morphology with the angular separation of a few $\times$ 0.1 pc. \citet{Seo15} observed this core as a part of a large-scale NH$_3$ survey with the Green Bank Telescope, cataloged as 12th core in their sample and derived a gas temperature of $\sim$9 K. \\
\ We have carried out ACA observations toward many starless cores including MC5-N (Sect. \ref{Obs:ACA}) targeting the dust continuum peaks obtained with single-dish telescopes, e.g., IRAM 30\,m (Sect. \ref{Obs:IRAM}). One of the main goals of this project is to understand the evolution of prestellar cores of low-mass stars by revealing their innermost (column) density structures (Sect. \ref{Intro:core}).  
The most concentrated dust continuum distribution has been unexpectedly found in the lowest mass condensation, MC5-N, among the observed samples collected with the ACA. 
We thus focus on MC5-N as a very evolved sub-solarmass condensation and a possible formation site of brown dwarf(s) or very low-mass star(s).
We have subsequently confirmed the high-deuterium fraction of this core using the N$_{2}$H$^{+}$ and N$_{2}$D$^{+}$ observations (Sect. \ref{Obs:45m}, \ref{R:N2HP}). In this paper, we present the physical/chemical properties of MC5-N with the angular resolution of $\sim$6$\arcsec$--20$\arcsec$ (Sect. \ref{Results}) and discuss the evolutionary state and formation process (Sect. \ref{Discussions}).

\section{Observations}\label{Obs.}
\subsection{1.2\,mm continuum: the IRAM 30\,m telescope}\label{Obs:IRAM}
\ The single-dish continuum observations of the 1.2mm thermal dust emission were performed between 2002 December and 2003 February with the IRAM 30\,m telescope on Pico Veleta (Spain) as a part of Project 109--02 (P.I.: Andr\'e, see also Sect. 2 of \citet{Tokuda16}), using the 117-channel MAMBO-2 bolometer camera (240 diameter) of the Max-Planck-Institute f\"{u}r Radioastronomie \citep{Kreysa99}.
The map contains a few additional dense core, MC5-S, MC6, and MC8 (see also Figure \ref{fig:1.2mm}(a)).
The FWHM  (full width at half-maximum) size of the main beam on the sky was measured to be 11$\arcsec$ based on beam maps of Uranus. The central effective frequency was 250 GHz, with halfsensitivity limits at 210 and 290 GHz. Pointing and focus positions were usually checked before and after each map. 

\subsection{1.2\,mm continuum: the Atacama Compact Array}\label{Obs:ACA}
\ We carried out ACA stand-alone mode observations in ALMA Cycle 4 (P.I.: Tachihara, \#2016.1.00928.S) in both Band 6 (1.2\,mm) dust continuum and molecular lines toward 16 dense cores located in nearby ($d$$\sim$150 pc) low-mass star-forming regions (Taurus, Ophiuchus-North, Chamaeleon, and Lupus). 
The ACA is composed of the 7\,m array as an interferometer and the Total Power (TP) array with the single-dish mode. The dust continuum and molecular line observations were conducted with the 7\,m array alone and with both the 7\,m array and the TP array, respectively. 
We selected the target cores based on the following criteria; (1) previous single dish observations confirmed relatively strong dust continuum emission (peak continuum flux $\gtrsim$\,30\,mJy\,beam$^{-1}$ with 14$\arcsec$beam at 1.2\,mm) compared to less evolved starless cores (the similar criterion was designed by \citet{Kirk05} as $``$bright$"$ cores), (2) cores are detected in high-density gas tracers, such as H$^{13}$CO$^{+}$ and N$_{2}$H$^{+}$, and (3) the previous infrared observations did not find any point-like sources therein. The observations have been done during October 2016 and May 2017. We selected a baseband containing two spectral windows, targeting H$^{13}$CO$^{+}$\,(3--2) and H$^{13}$CN\,(3--2), and the correlator for each spectral window was set to have a bandwidth of 58.59\,MHz with 1920 channels. We used three basebands for the continuum emission with a bandwidth of 1875\,MHz. The $uv$ range of the 7\,m array data is 6.5--34\,k$\lambda$. We performed the data reduction using the Common Astronomy Software Application (CASA) package \citep{McMullin07} version 5.0.0. We used the natural weighting for the CLEAN process and the resultant synthesized beam size of the MC5-N observation was 7$\farcs$5 $\times$ 5$\farcs$2 (PA = $-$55$\degree$), which is corresponding to $\sim$1050 au $\times$ 840 au at a distance of Taurus ($\sim$140 pc, \cite{Elias78}). The sensitivity of the continuum observation (1$\sigma$ noise level) was 0.8\,mJy\,beam$^{-1}$. \\
\ In this paper, we only describe the results in MC5-N, which has the strongest compact emission among the observed samples in the survey. The full analysis encompassing all the sources in both the continuum and the molecular lines will be presented in a forthcoming paper. 

\subsection{N$_{2}$H$^{+}$ and N$_{2}$D$^{+}$: the Nobeyama 45 m telescope}\label{Obs:45m}
\ We carried out N$_{2}$H$^{+}$(1--0) observations of $\sim$40 dense cores in the Taurus molecular cloud with the Nobeyama 45 m telescope between February 2006 and March 2007 (P.I.: T., Onishi). The target selection was based on the early H$^{13}$CO$^+$ survey in this region performed by \citet{Onishi02}. Our present target, MC5 was observed in N$_{2}$H$^{+}$ (1--0) as a part of this survey. The observation mode was the position switching using the multi-beam (25 beam) dual-sideband SIS receiver, BEARS \citep{Sunada00}. The FWHM beam size and the grid-spacing were 17$\farcs$8 and 20$\farcs$55, respectively. The typical system noise temperature was $\sim$200 K, and the resultant noise level was $\sim$0.15\,K in the $T_{\mathrm A}^*$ scale at a velocity resolution of $\sim$0.1 km\,s$^{-1}$. The main-beam efficiency at 93\,GHz was $\sim$50\%. The detailed descriptions of the observations and the full analysis of all the sources will be shown in a forthcoming paper (Miyamoto et al. in prep.).\\
\ The N$_{2}$D$^{+}$\,(1--0) observations were carried out in the on-the-fly (OTF) mapping mode \citep{Sawada08} using the T70 dual-polarization and two-sideband SIS receiver on 2017 December 18th. Although our main target line is N$_{2}$D$^{+}$ (1--0), we simultaneously observed the other molecular lines, HCO$^{+}$\,(1--0), HNC\,(1--0), and DNC\,(1--0). The data sampling interval along a strip and separations between the strips are 6$\arcsec$. The effective beam size was $\sim$23$\arcsec$. The telescope pointing was established by observing an SiO maser source, NML Tau, at 43\,GHz every $\sim$1 hour during the observations. The achieved pointing accuracy was a few arcseconds. The system noise temperature was 160--200\,K during the observations and the resultant noise level was $\sim$0.1\,K in the $T_{\mathrm A}^*$ scale at a velocity resolution of $\sim$0.1 km\,s$^{-1}$. The main-beam efficiency at 77\,GHz was 54\%. 

\section{Results}\label{Results}
\subsection{1.2\,mm dust continuum observations with the IRAM 30\,m telescope and the ACA}\label{R:1.2mm}
\begin{figure*}[t]
 \begin{center}
  \includegraphics[width=14cm]{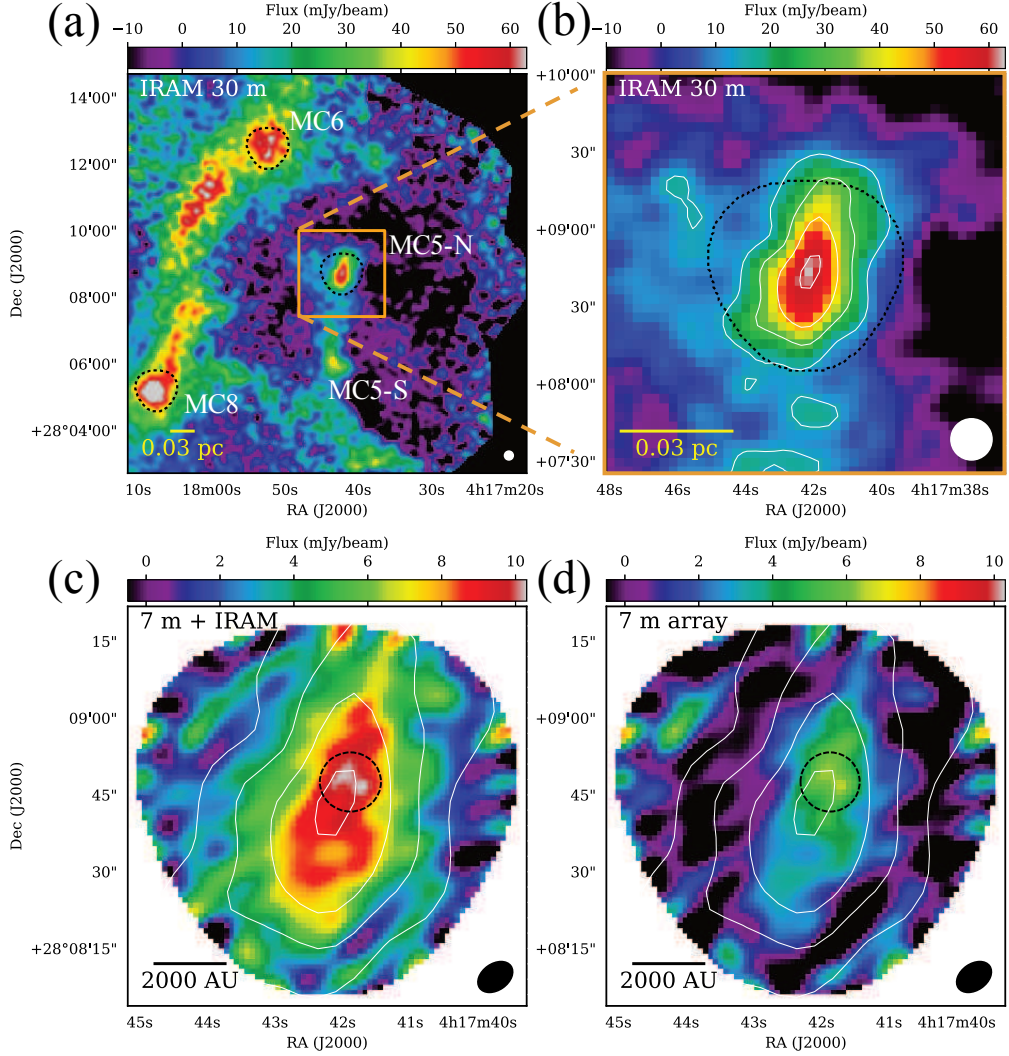}
 \end{center}
 \caption{Distribution of 1.2\,mm dust continuum emission toward MC5-N. (a) Color scale image shows 1.2\,mm dust emission obtained by the IRAM 30\,m telescope. The dotted lines show the field coverage of the 7\,m array observation toward MC5-N, MC6 and MC8. (b) An enlargement view of the orange rectangle in panel (a). The lowest contour and subsequent contour step are \textcolor{black}{15\,mJy\,beam$^{-1}$}. The dotted line shows the field coverage of the 7\,m array observation shown in panel (c,d). The angular resolution is given in the lower right corner, 16$\arcsec$. 
(c) Same as panel (b) but for the combined (7\,m array + IRAM 30\,m) data shown in color scale. (d) Same as panel (b) but for the 7\,m array data alone. The angular resolution is given in the lower right corners, 7$\farcs$3 $\times$ 5$\farcs$0 in panel (c,d). \textcolor{black}{The dotted circles in panel (c,d) indicate the positions of the possible substructures.}}\label{fig:1.2mm}
\end{figure*}

\begin{figure*}
 \begin{center}
  \includegraphics[width=16cm]{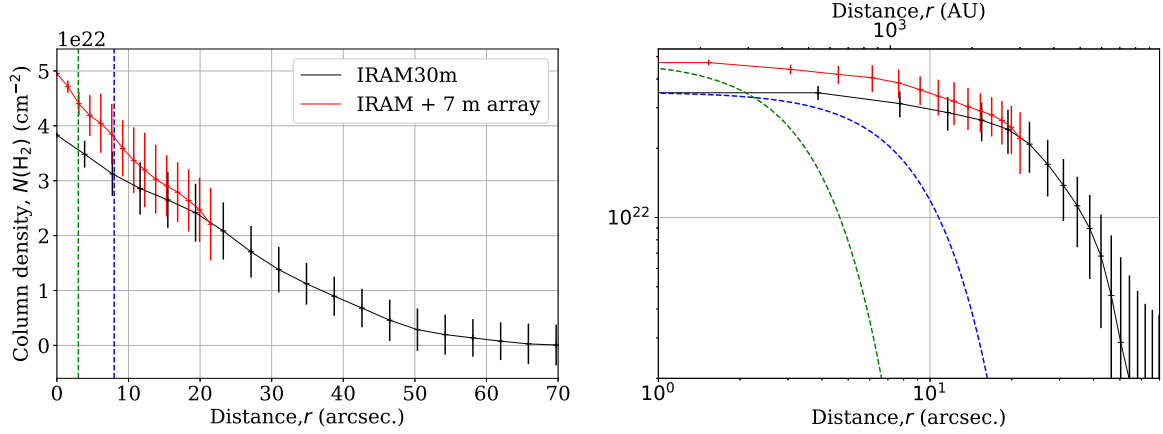}
 \end{center}
 \caption{
Mean radial profiles of H$_2$ column density centered at the peak position in MC5-N derived from the 1.2\,mm dust continuum image obtained by the IRAM 30\,m telescope alone and the combined 7\,m array + IRAM data. Left and right panels show the linear--linear plot and the log--log plot of the profiles, respectively. The averaged profiles of the combined data and the IRAM data are shown by black and red solid lines, respectively. Black and red bars show the ($\pm$1$\sigma$) dispersion of the distribution of the radial profiles in each data. 
Green/blue lines in the left panel shows the half widths at half maximum of the beams, \textcolor{black}{8}$\arcsec$ and 3$\arcsec$. In the right panel green/blue curves are same as in the left panel but for Gaussian functions with the same widths.}\label{fig:profile}
\end{figure*}

\ Figure \ref{fig:1.2mm} shows the 1.2\,mm continuum distributions of MC5-N obtained by the IRAM 30\,m/MAMBO-2 and the 7\,m array of the ACA (hereafter, the 7\,m array). Panels (a) and (b) show the large-scale view and the enlarged view, respectively, obtained with the IRAM 30\,m alone. In panel (a), there are two major filamentary structures containing several local peaks (i.e., condensations), MC5-N/S, MC6 and MC8, which are also identified by the previous molecular line and dust observations (e.g., \cite{Onishi02,Hacar13,Tafalla15,Seo15,Buckle15,Ward-Thompson16}). 
As shown in panel (a), the width and intensity of the eastern side filamentary cloud associated with MC6 and MC8 are much wider/higher than those of the western one containing MC5-N/S. \\
\ We carried out the 7\,m array observations toward the three condensations, MC5-N, MC6 and MC8, shown in Panel (a) as a parts of our project (Sect. \ref{Obs:ACA}).
Panels (c) and (d) show the combined (7\,m + IRAM) 1.2\,mm data and the 7\,m array data alone toward MC5-N. We used the feathering technique to combine the two different data sets in accordance with the CASA instruction (see also \cite{Tokuda16}). The single-dish dust continuum emission in the IRAM image well correlates with that in the 7\,m array alone (Figure \ref{fig:1.2mm}(d)). We found that the peak position of the 7\,m array image is slightly shifted ($\sim$6$\arcsec$) from that of the IRAM image. The 0.87\,mm continuum image obtained with the JCMT/SCUBA2 \citep{Buckle15,Ward-Thompson16} and a high-density gas tracer, N$_{2}$D$^{+}$ (1--0), obtained with the Nobeyama 45 m (see Sect. \ref{R:N2HP}) show the same peak position in the IRAM image. 
A similar positional discrepancy between the 7\,m array and a single-dish observation is also reported in another starless core, L1689N \citep{Lis16}. We note that the 7\,m array observation revealed that the core center is marginally resolved into two peaks \textcolor{black}{as shown by the dotted circles in Figure \ref{fig:1.2mm} (c,d)}. \textcolor{black}{The intensity difference between the peaks and the local minimum is less than 1\,$\sigma$ noise level, and the separation is $\sim$4$\arcsec$.  We need higher angular resolution observations to confirm the substructures, and the possible origins of the substructures will be discussed in Sect. \ref{D:2}.}\\
We derived the column density and the mass from the 1.2\,mm continuum emission to evaluate the physical properties of the core assuming uniform dust temperature $T_{\rm d}$, and the optically thin emission. The H$_2$ column density, $N$(H$_2$), is estimated using the following equations,
\begin{equation}
N_\mathrm{H_2} = \frac{F_\nu^\mathrm{beam}} {\Omega_\mathrm{A}\mu_\mathrm{H_2} m_\mathrm{H} \kappa_\nu B_\nu (T_\mathrm{d})}
\label{eq:NH2} 
\end{equation}
and
\begin{equation}
\kappa_\nu = \kappa_\mathrm{231GHz} (\nu/\mathrm{231GHz})^{\beta},
\end{equation}
where $F_\nu^\mathrm{beam}$ is the flux per beam at frequency $\nu$, $\Omega_\mathrm{A}$ is the solid angle of the beam, $\mu_{\rm H_2}$ is the molecular weight per molecular hydrogen,
$m_{\rm H}$ is the H-atom mass, $\kappa_\nu$ is the mass absorption coefficient, $\kappa_\mathrm{231 GHz}$
is the emissivity of the dust continuum emission at 231\,GHz 
$\beta$ is the dust emissivity index, $B_\nu$ is the Plank function. We adopt a $\beta$ of 2 (c.f., \cite{Ormel07}).
\textcolor{black}{Based on the NH$_{3}$ observations, \citet{Seo15} derived the gas temperature of this core as $\sim$9 K, and
and this is consistent with the dust temperature of $\sim$8\,K derived by the {\it Herschel} Gould Belt survey \citep{Marsh16}.
We use 9\,K as the dust temperature to derive the column density in this study. }
\textcolor{black}{The uncertainty of column density estimation is caused by the assumption of the dust opacity and the temperature distribution of the inner part. 
The dust opacity, $\kappa_\mathrm{231 GHz}$, can change in the range of 0.02--0.005\,cm$^2$\,g$^{-1}$ (see, e.g., \cite{Oss94,Motte98,Kauffmann08}), which changes the column density estimation linearly as in Equation (\ref{eq:NH2}). \textcolor{black}{For example, $\kappa_\mathrm{231 GHz}$ = 0.005\,cm$^2$\,g$^{-1}$ is used for prestellar cores (e.g., \cite{Preibisch93}), $\kappa_\mathrm{231 GHz}$ = 0.01\,cm$^2$\,g$^{-1}$ is used for circumstellar envelopes around Class I and Class 0 protostars (e.g., \cite{Oss94}). Taking into account the evolved nature of this core (see also Sect. \ref{R:N2HP}), the evolutionary state of this core is considered to be in between prestellar and protostellar phases. Therefore, we consider two values, $\kappa_\mathrm{231 GHz}$ = 0.005 and 0.01\,cm$^2$\,g$^{-1}$ to derive the column density (see also Table \ref{table:1.2mm_prop}).}
Although we assume an isothermal temperature distribution to derive the total mass, the gas/dust temperature profile may significantly drop toward the center (e.g., \cite{Evans01}). For example, based on the Very Large Array (VLA) observations in NH$_3$, \citet{Crapsi07} reported a gas temperature of $\sim$6\,K toward the center of an evolved starless core, L1544, which is in a similar evolutionary \textcolor{black}{stage as} MC5-N (see Sect. \ref{R:N2HP}). A similar temperature profile can thus be expected for MC5-N at the center, and the (column) density estimation of the central part may increase depending on Equation (\ref{eq:NH2}).} \\
\ Although there is no convincing method to measure the elongations of dense cores along the line-of-sight, a circularly averaged column density profile can be a good indicator of the density enhancement and evolutionary state of dense cores (e.g., \cite{Ward94,Ward99,Motte01,Kandori05,Kirk05,Tokuda16}). Since the shape of MC5-N is elongated along the north-south direction, we performed 2D Gaussian fitting to the column density distribution obtained from the IRAM 30\,m data alone, and the $``$equivalent$"$ radius $r = (ab)^{1/2}$ was substituted for the radius, where $a$ and $b$ are the semi-major and semi-minor axes of the ellipse of the core, respectively. The fitted parameters, the major/minor axes in FWHM and position angle (PA), which is measured counter clockwise relative to the north celestial pole, are shown in Table \ref{table:1.2mm_prop}. We manually masked the elongated features toward the south and east directions, corresponding to the fan-shaped area within PA = 50--250$\degree$ centered on the peak. The resultant averaged column density profiles obtained from the 1.2\,mm continuum data \textcolor{black}{assuming $\kappa_\mathrm{231 GHz}$ = 0.005\,cm$^2$\,g$^{-1}$} are shown in Figure \ref{fig:profile}. 
While the IRAM data alone cannot fully trace the inner density profile due to the beam dilution, the 7\,m array combined data clearly resolved the inner (column) density enhancement. \\
\ To estimate the averaged properties around the central position, we estimated the radius of the inner flattened region, $R_{\rm flat}$, \textcolor{black}{using} Plummer-like function fitting to the column density profile (c.f., \cite{Ward99,Tafalla02}) \textcolor{black}{with three parameters; $R_{\rm flat}$, $\average{N_{\rm H_2}}_{R_{\rm flat}}$, and the asymptotic power index.} 
We summarize the physical properties of MC5-N derived from the 1.2\,mm continuum data in Table \ref{table:1.2mm_prop}.
\textcolor{black}{We estimated the mass enclosed in a half-intensity contour, corresponding to the diameter of 6.8\,$\times$\,10$^3$\,au (=0.03\,pc), to be $M_{\rm FWHM}$ $\sim$0.2--0.4\,$M_{\odot}$.  The central density ($\average{n_{\rm H_2}}_{R_{\rm flat}}$), and averaged density ($n_{\rm ave}$) are $\sim$(1--2)\,$\times$\,10$^{6}$\,cm$^{-3}$, and $\sim$(4--7)\,$\times$\,10$^5$\,cm$^{-3}$, respectively. The average density of this core is not significantly different from those of typical prestellar cores in Taurus (c.f., \cite{Onishi02}), and the central density, however, is significantly higher than the averaged density. This result implies that the core has a steep density profile.} \textcolor{black}{For example, the total mass within the 10\% intensity contour of the peak flux of the IRAM data ($M_{\rm 10\%}$) has only a slight higher value of 0.3--0.6\,$M_{\odot}$. It should be noted that the ground-based continuum observations by the IRAM 30\,m telescope suffer from a progressive loss of signal as angular radius increases \citep{Motte01} and the MC5-N core is located in the high-column density background in the L1495 region. Therefore, the column density is probably underestimated at the outer radii. We compared our column density map with that derived from the {\it Herschel} observations to estimate this effect. The $M_{\rm FWHM}$ is not significantly different in both data: actually, \citet{Marsh16} using the {\it Herschel} data also derived the similar mass of $\sim$0.4\,$M_{\odot}$. However, the $M_{\rm 10\%}$ estimated form the {\it Herschel} column density map is factor of $\sim$2 higher than that of the IRAM data. }
\\
\ We briefly mention comparisons between MC5-N and the other objects obtained in our ACA survey. Despite MC5-N \textcolor{black}{having} the lowest intensity of single-dish based 1.2\,mm emission among our present samples (16 targets in total), it has the highest intensity and the most concentrated distributions in the 7\,m array observations. In the case of MC6 and MC8 shown in Figure \ref{fig:1.2mm} (a), we did not detect any significant emissions in 1.2\,mm continuum with the 7\,m array. 
These results show that MC5-N is the most evolved core (i.e., the densest core) prior to protostar formation in our sample and the previous single-dish observations in dust continuum cannot fully trace the density enhancement at the innermost regions of starless cores.

\begin{table*}
\tbl{Physical properties of MC5-N derived from the 1.2\,mm emission.}{%
\begin{tabular}{*{11}{c}}  
\hline\noalign{\vskip3pt}
$N_\mathrm{peak}^\mathrm{IRAM}$\,$^{a}$ & $N_\mathrm{peak}^\mathrm{7\,m}$\,$^{b}$ & $R_{\rm flat}$\,$^{c}$ 
& $\average{N_{\rm H_2}}_{R_{\rm flat}}$\,$^{d}$ & $\average{n_{\rm H_2}}_{R_{\rm flat}}$\,$^{e}$ & Mass\,$^{f}$ & Major axis\,$^{g}$ & Minor axis\,$^{g}$ & PA\,$^{g}$ & Diameter\,$^{h}$ & \textcolor{black}{$n_{\rm ave}$\,$^{i}$} \\
(cm$^{-2}$)& (cm$^{-2}$) & (au) & (cm$^{-2}$) & (cm$^{-3}$) & ($M_{\odot}$) & (au) & (au) & (degree) &(au) & \textcolor{black}{(cm$^{-3}$)}\\
\hline\noalign{\vskip3pt} 
\textcolor{black}{(1.7--3.5)}\,$\times$\,10$^{22}$ & \textcolor{black}{(2.4--4.8)}\,$\times$\,10$^{22}$ & 1.0\,$\times$\,10$^3$ 
& \textcolor{black}{(1.8--3.6)}\,$\times$\,10$^{22}$ & \textcolor{black}{(1--2)}\,$\times$\,10$^{6}$ & \textcolor{black}{0.2--0.4} & 9.2\,$\times$\,10$^3$ & 5.0\,$\times$\,10$^3$   & 159  & 6.8\,$\times$\,10$^3$ & \textcolor{black}{\textcolor{black}{(4--7)}\,$\times$\,10$^5$}\\
\hline\noalign{\vskip3pt} 
\end{tabular}}\label{table:1.2mm_prop}
\begin{tabnote}
{\footnotemark[$a$]\hss}\unskip%
The peak column density measured in a $\sim$11$\arcsec$ beam centered on the source.\\
{\footnotemark[$b$]\hss}\unskip%
The peak column density measured in a $\sim$6$\arcsec$ beam centered on the source.\\
{\footnotemark[$c$]\hss}\unskip%
The flat inner radius, $R_{\rm flat}$, is determined by fitting the column density profile of the combined (7m + IRAM) data with a Plummer-like function.\\
{\footnotemark[$d,e$]\hss}\unskip%
The column and volume densities averaged over the core inner flat region, $R_{\rm flat}$. \\
{\footnotemark[$f$]\hss}\unskip%
The total mass integrated above the half intensity of the peak continuum emission.\\
{\footnotemark[$g$]\hss}\unskip%
The derived parameters by 2D gaussian fitting to the column density distributions obtained from the IRAM 30\,m data alone (see the text).\\
{\footnotemark[$h$]\hss}\unskip%
The geometric mean between the major and minor axis.\\
{\footnotemark[$i$]\hss}\unskip%
\textcolor{black}{The averaged density calculated from the total mass and diameter with an assumption of the spherical geometry.}\\

\end{tabnote}
\end{table*}


\subsection{N$_{2}$H$^{+}$ and N$_{2}$D$^{+}$ observations with the Nobeyama 45 m telescope}\label{R:N2HP}
\begin{figure*}[t]
 \begin{center}
  \includegraphics[width=16cm]{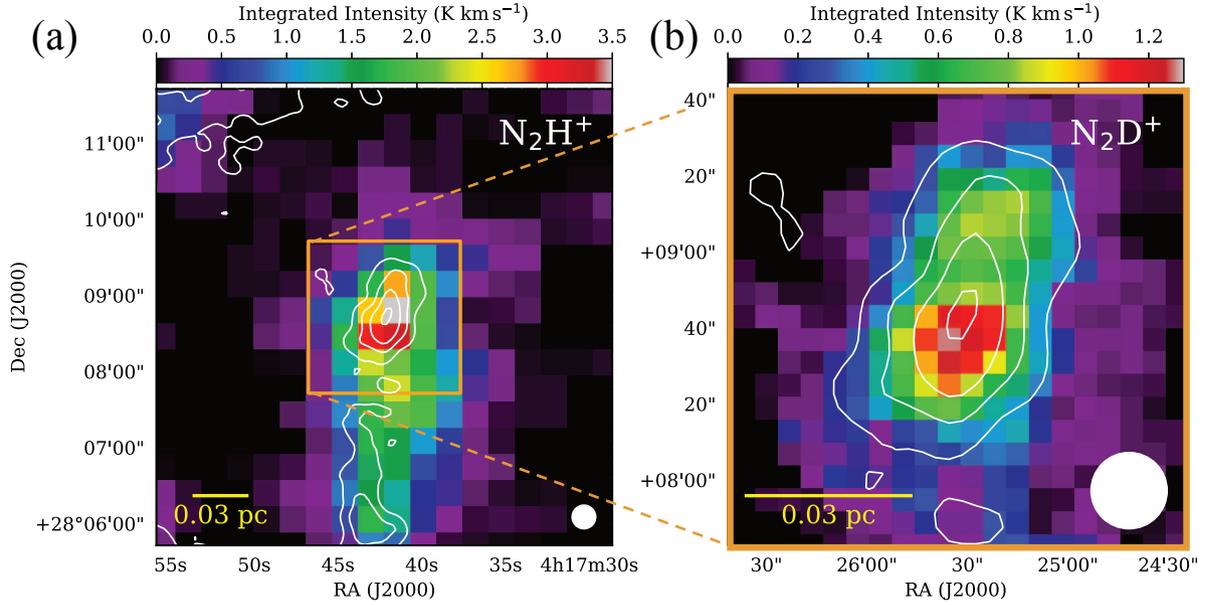}
 \end{center}
 \caption{Integrated Intensity images of N$_{2}$H$^{+}$ (1--0), and N$_{2}$D$^{+}$ (1--0) in MC5-N. The contours in both panels show the 1.2\,mm continuum emission obtained with the IRAM 30\,m. The contour levels are same as Figures 1. The angular resolution is shown in the lower right corners in each panel.}
 \label{fig:iiN2HP}
\end{figure*}

\begin{figure}
 \begin{center}
  \includegraphics[width=8cm]{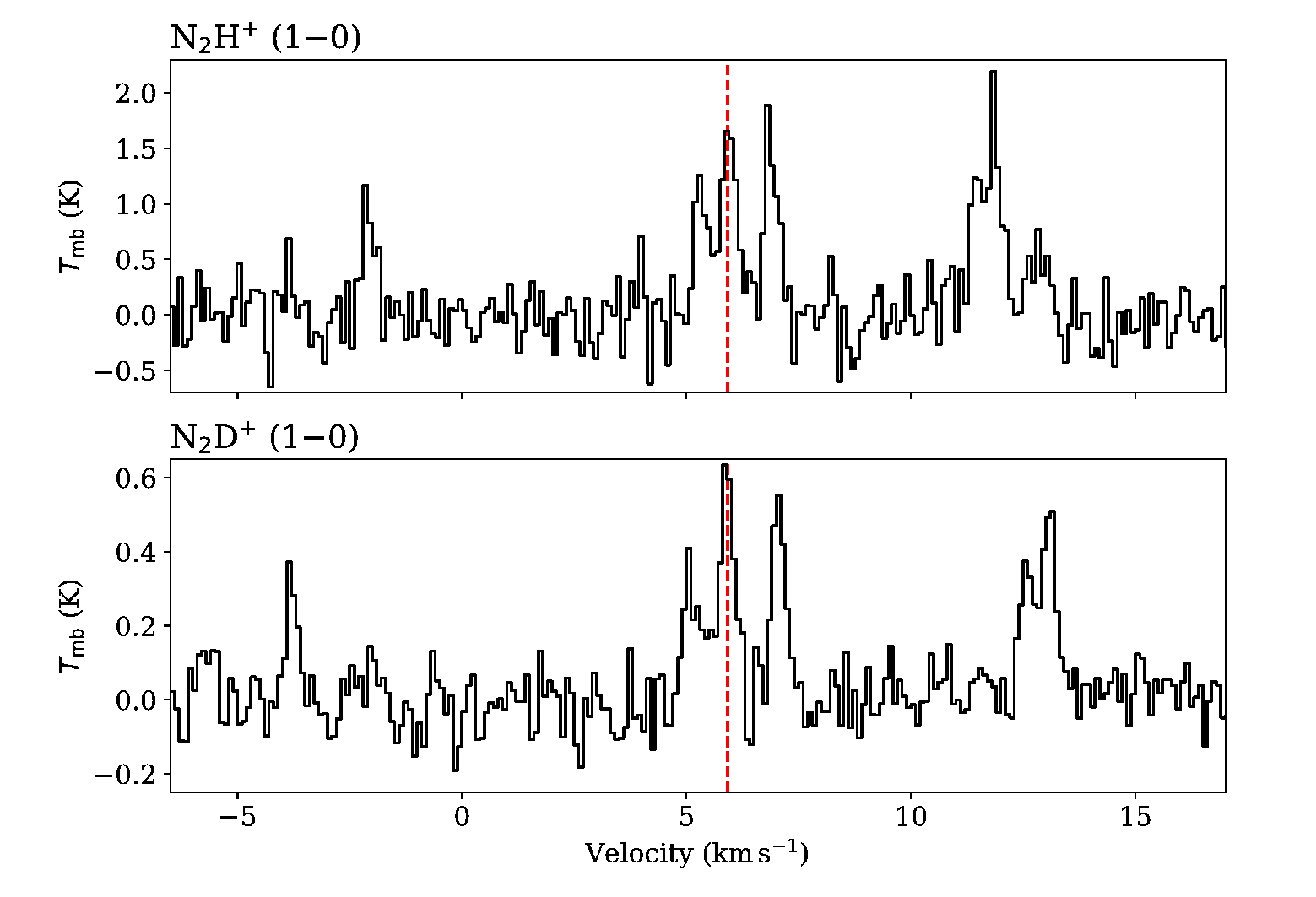}
 \end{center}
 \caption{N$_2$H$^+$ (1--0), and N$_2$D$^+$ (1--0) spectra toward the dust continuum peak of MC5-N. Red vertical lines show the centroid velocity, 5.92 km\,s$^{-1}$, deduced from the hyperfine fitting of the N$_2$H$^+$ spectrum.}
 \label{fig:spectN2HP}
\end{figure}

\ Figures \ref{fig:iiN2HP} and \ref{fig:spectN2HP} show integrated intensity images and spectra, respectively, of N$_{2}$H$^{+}$ (1--0) and N$_{2}$D$^{+}$ (1--0) toward MC5-N obtained with the Nobeyama 45 m telescope. The spatial distributions of the two molecules are roughly correlated with the 1.2\,mm dust continuum image, indicating that they are well tracing the column density of the core. 
We also checked a fully sampled archival N$_{2}$H$^{+}$ (1--0) data toward this core obtained by the IRAM 30\,m \citep{Tafalla15} with a beam size of $\sim$33$\arcsec$, we cannot find a significant difference in the intensity distributions between the N$_{2}$H$^{+}$ and the 1.2\,mm emissions. \\
\ We estimated the physical quantities, e.g., the column density and velocity dispersion, to evaluate the evolutionary state of the core. We derived the deuterium fraction, which is considered to increase monotonically until the stage of protostar formation, as an indicator of the evolutionary state of starless cores. We fit the observed spectra at the center of the core using the hyperfine spectrum model of multiple Gaussian components with the line optical depth effect (c.f., \cite{Tine00,Tatematsu04}). The details of the column density estimation are given in \citet{Caselli02}. The estimated properties are listed in Table \ref{table:coreN2HP}. The properties of MC5-N resemble to those of other starless cores in low-mass star-forming regions (e.g., \cite{Tatematsu04,Crapsi05,Daniel07}). One of the striking features is the high-column density ratio, $N$(N$_{2}$D$^{+}$)/$N$(N$_{2}$H$^{+}$), $\sim$\textcolor{black}{30\%}. This fact indicates that the deuterium fractionation is well developed in the high-density region. Such a high-column density ratio is rarely seen in other cores except for extremely evolved prestellar core on the verge of star formation, e.g., L1544 \citep{Caselli02,Crapsi05} and a first core candidate, B1b-N \citep{Hung13}.\\
\ To understand the dynamical state of the core, we estimated the non-thermal velocity components using the following equation,
\begin{equation}\label{eq:nonT}
\Delta v^2_{\rm NT} = \Delta v_{\rm deconv}^2 - 8\,\mathrm{ln}\,2\,k_{\rm B}T_k/m_{\rm obs},
\end{equation}
where $k_{\rm B}$ is the Boltzmann constant, $T_k$ is the kinetic temperature, $\Delta v_{\rm deconv}$ is the observed linewidth (FWHM) deconvolved with the velocity resolution of $\sim$0.1 km\,s$^{-1}$, and $m_{\rm obs}$ is the mass of the observed molecule. If we adapted the $T_{k}$ of \textcolor{black}{9}\,K based on the NH$_{3}$ observations \citep{Seo15}, the $\Delta v_{\rm NT}$ of N$_{2}$D$^{+}$ is calculated to be 0.29 km\,s$^{-1}$. \textcolor{black}{This is consistent with the non-thermal velocity dispersion derived from the NH$_{3}$ observation, $\sigma_{\rm NT}$ $\sim$0.1--0.15\,km\,s$^{-1}$ (Figure 8 in \citet{Seo15}).}
Since the thermal linewidth, $\Delta v_{\rm T}$, corresponding to the second term on the right hand of the equation (\ref{eq:nonT}), for a mean particle mass of 2.33 amu \textcolor{black}{($m_{\rm mean}$)} is  0.44 km\,s$^{-1}$, the dynamical state of this core is considered to be dominated by thermal motion rather than turbulence.
\textcolor{black}{We derived the virial mass, $M_{\rm vir}$ using the correction coefficient to use for the potential energy term described by \citet{Sipila11}. The effective linewidth, $\Delta v_{\rm eff}$, for the virial mass calculation is defined by the following equation, 
\begin{equation}\label{eq:Veff}
\Delta v^2_{\rm eff} = \Delta v_{\rm deconv}^2 + 8\, \mathrm{ln}\, 2\, k_{\rm B}T_k(\frac{1}{m}-\frac{1}{m_{\rm obs}}).
\end{equation}
The averaged $\Delta v_{\rm deconv}$ of N$_2$D$^{+}$ is measured to be $\sim$0.26\,km\,s$^{-1}$ within the core radius of 0.015\,pc. This is slightly smaller than the linewidth at the center position (Table \ref{table:coreN2HP}). If we assume a kinematic temperature, $T_k$, of 9\,K, the resultant $M_{\rm vir}$ is $\sim$0.6\,$M_{\odot}$. In this case, \textcolor{black}{the} virial parameter, $\alpha_{\rm vir}$ = $M_{\rm vir}$/$M$, where $M$ is the mass derived by the dust continuum observations (Sect. \ref{R:1.2mm}), is $\sim$3--1 depending on the assumption of the opacity.
Therefore, the virial parameter is estimated to be order of unity, and the core is seems to be gravitationally bound.} Based on the evolved nature of this core, the steep column density profile, the high deuterium fractionation and the narrow linewidth, we conclude that MC5-N is one of the most evolved cores \textcolor{black}{in the Taurus molecular cloud}. The core mass derived from the dust emission, $\sim$\textcolor{black}{0.2--0.4}\,$M_{\odot}$ (Sect. \ref{R:1.2mm}) is an order of magnitude smaller than ordinary prestellar cores. We discuss the possibility that MC5-N is a unique source as a prestellar core of very low-mass objects, such as brown dwarf(s) or very low-mass star(s), in Sect. \ref{Discussions}.
\\

\begin{table}
\tbl{Physical properties at the center of MC5-N derived from the N$_{2}$H$^{+}$/N$_{2}$D$^{+}$ observations.}{%
\begin{tabular}{*{7}{c}}  
\hline\noalign{\vskip3pt} 
Molecule & $V_{\rm lsr}$  & $\Delta v_{\rm obs}$\,$^{a}$ 
& $T_{\rm ex}$\,$^{b}$ & $\tau_{\rm total}$\,$^{c}$ & $N_{\rm tot}$\,$^{d}$ & $\Delta v_{\rm NT}$\\
& (km\,s$^{-1}$) & (km\,s$^{-1}$)
& (K) &  & (cm$^{-2}$)  & (km\,s$^{-1}$)\\
\hline\noalign{\vskip3pt} 
N$_2$H$^+$ & 5.92 & 0.34 & 5.3         & 5.9 & 6.3 $\times$ 10$^{12}$ & 0.30\\
N$_2$D$^+$ & 5.91 & 0.33 & $\cdots$ & 1.6 & \textcolor{black}{1.9} $\times$ 10$^{12}$ & 0.29\\
\hline\noalign{\vskip3pt} 
\end{tabular}}\label{table:coreN2HP}
\begin{tabnote}
{\footnotemark[$a$]\hss}\unskip%
 Observed linewidth (FWHM) of the spectrum derived by the hyper fine fitting.\\
{\footnotemark[$b$]\hss}\unskip%
 Excitation temperature. The fixed value obtained from the N$_2$H$^+$ analysis is applied in the N$_2$D$^+$ fitting to derive the other parameters.\\
{\footnotemark[$c$]\hss}\unskip%
Sum of the optical depths of the hyper fine components. \\
{\footnotemark[$d$]\hss}\unskip%
 Column density of the molecule.\\
\end{tabnote}
\end{table}


\section{Discussions}\label{Discussions}
\subsection{A candidate of a prestellar core forming brown dwarf(s)?}\label{D:1}
\ The sub-stellar-mass condensation, MC5-N, shows high density ($\sim$10$^{6}$\,cm$^{-3}$), a high deuterium fraction ($\sim$\textcolor{black}{30\%}), and narrow linewidth close to the thermal motion at a temperature of 10 K, indicating that this core is in a well developed stage in the dense core evolution.
\textcolor{black}{As shown in Sect. \ref{R:N2HP}, this core is considered to be gravitationally bound. Since the density of the outer edge of this core is $\sim$10$^4$\,cm$^{-3}$ based on the C$^{18}$O observations \citep{Onishi96,Onishi98,Hacar13,Tafalla15}, the density contrast between the center and the edge is $\sim$100.  If we consider the Bonner-Ebert sphere model \citep{Bonner56,Ebert55}, the density contrast of $>$14 means that the equilibrium state is unstable against gravitational collapse \citep{Kandori05,Kirk05}. 
If we adopt a magnetic field strength $\sim$25\,$\mu$G derived using the Chandrasekhar-Fermi method for the large-scale infrared polarization observation \citep{Chapman11}, the critical mass is calculated to be $\sim$0.05\,$M_{\odot}$ for a core with the diameter of 0.03\,pc and the magnetic field orientation is roughly parallel to the core elongation. These facts may indicate that the magnetic field does not help to stabilize the core.} \\
\ The deuterium fraction ($D_{\rm frac}$) deduced from the $N$(N$_{2}$D$^{+}$)/$N$(N$_{2}$H$^{+}$) ratio may provide us with a timescale (i.e., evolutionary state) of the dense core. Typical starless cores with a density of $\sim$10$^{5}$\,cm$^{-3}$ and the flat inner density profile (e.g., B65, \cite{Alves01}) show that the $D_{\rm frac}$ estimated from the N$_2$H$^+$ and N$_2$D$^+$ observations is lower than 0.1 \citep{Crapsi05}. If the $D_{\rm frac}$ monotonically increases as a function of time, MC5-N should be in a later evolutionary stage.
According to the recent numerical calculations including a complete chemical network with the initial H$_2$ density of 10$^{5}$\,cm$^{-3}$ \citep{Kong15}, it takes $\sim$10$^{6}$ yr to reach the $D_{\rm frac}$ of $\sim$0.2 independently on the initial ortho-to-para H$_2$ ratio, which is consistent with that of MC5-N in this study (Sect. \ref{R:N2HP}). 
On the other hand, other chemical network calculations suggest that high deuteration fractions of the order $\sim$0.1 may be reached within two free-fall times with a density of $\sim$10$^{5}$\,cm$^{-3}$ by considering more realistic initial conditions, such as filamentary configurations with magnetic fields and turbulence \citep{Kortgen18}.
This shorter timescale is consistent with the lifetime of typical starless cores with densities of 10$^{5}$\,cm$^{-3}$, several $\times$ 10$^{5}$ yr, derived by statistical studies based on a large number of samples \citep{Onishi02,Kirk05,Ward07}.
Although it is hard to estimate the exact timescale of the core from the observed $D_{\rm frac}$ so far, the current observational/theoretical results in the literature suggest that MC5-N should be at the last stage of the starless phase.\\
\ If we assume a typical \textcolor{black}{core to} star formation efficiency of $\sim$20\%--40\% in nearby molecular clouds \citep{Andre10}, the stellar object(s) that will be formed in this core is considered to be a brown dwarf(s) or a very low-mass star(s).
Numerical simulations by \citet{Machida09} show a brown dwarf mass object is formed in a rotating magnetized compact cloud with a mass of 0.22\,$M_{\odot}$ in the same manner as hydrogen-burning stars. The star formation efficiency is suppressed as low as 20\% since a large portion of the parental cloud mass is ejected by the protostellar outflow (see also \cite{Machida13}). 
Our results as well as the theoretical studies suggest that a brown dwarf is formed from a sub-stellar mass condensation via gravitational collapse. 
\textcolor{black}{We note a few causes of uncertainty regarding the mass of the forming protostars.} 
\textcolor{black}{\citet{Marsh16} suggested that the core to star formation efficiency in the L1495 region can be higher than that of the Aquila region \citep{Andre10} by a factor of two, which affects the mass estimation of the formed star, although they indicated that the difference in the efficiencies in the two regions may be within the uncertainties.
It also should be noted that the above estimates assume that the mass of the core does not change during the prestellar/protostellar collapse phase. 
If the core gains further materials as it evolves, the forming protostar mass could be more massive.} 
\\
\subsection{Formation of the \textcolor{black}{very low-}mass condensation}\label{D:2}
\ Next we discuss origins of the \textcolor{black}{very low-}mass condensation. 
In the low-mass star-forming Taurus molecular cloud, we exclude the possibility that the core became smaller in the photo-erosion processes driven by massive stars. We here consider two mechanisms as the origin of MC5-N, (1) fragmentation of a filamentary molecular cloud, and (2) shock compression and fragmentation driven by turbulent motions. \\
\ Recent {\it Herschel} surveys (\cite{Andre14} and references therein) in nearby molecular clouds suggest that majority of low-mass prestellar cores are formed in $``$supercritical$"$ filaments whose mass exceed the critical line mass, $M_{\rm line,crit}$ = 2$c_{\rm s}^2$/G (e.g., \cite{Stod63,Ostriker64,Inutsuka92}), where $c_{\rm s}$ $\sim$0.2 km\,s$^{-1}$ is the isothermal sound speed for a gas temperature of $\sim$10 K. They also found a column density threshold for prestellar cores, $N$(H$_2$) $\sim$ 7 $\times$ 10$^{21}$\,cm$^{-2}$ (see also \cite{Onishi98}), which roughly corresponds to the core boundary of MC5-N judging from the N$_2$H$^{+}$, N$_2$D$^{+}$, and 1.2\,mm observations.  
The core diameter and mass are 0.03 pc and \textcolor{black}{0.2--0.4}\,$M_{\odot}$, respectively (Table \ref{table:1.2mm_prop}), which are consistent with the thermal Jeans length and mass with the gas density of $\sim$10$^{6}$\,cm$^{-3}$ at a gas temperature of $\sim$10 K.  Although typical starless cores are considered to be formed by the gravitational instability of less dense ($\sim$10$^{4}$\,cm$^{-3}$) clouds (e.g., \cite{Onishi02}), the MC5-N condensation may be a fragment of a much higher density filament after the radial direction collapse as suggested by \authorcite{Inutsuka92} (\yearcite{Inutsuka92,Inutsuka97}). In fact, \textcolor{black}{MC5-N is indeed located in a narrow filamentary structure based on the continuum map shown in Figure \ref{fig:1.2mm} (a), and another sub-core, MC5-S, is also located in the same filament, implying} that there were fragmentation processes in this system. \textcolor{black}{We estimated the physical parameters of MC5-S using the same methods described in Sect. \ref{R:1.2mm}. 
The size of MC5-S is similar to that of MC5-N and the total mass and average density are $\sim$0.2--0.3\,$M_{\odot}$ and $\sim$(1--3)\,$\times$\,10$^{5}$\,cm$^{-3}$, respectively. In addition to this, our ongoing survey using the Nobeyama 45\,m telescope has detected the N$_2$D$^{+}$\,(1--0) emission toward the dust continuum peak of MC5-S. The intensity is similar to that of MC5-N, indicating that the central density may not be significantly different between the two sources.} The width of the filament along the MC5-N/S cores is much narrower than that of another filament containing the MC6,8 cores shown in Figure \ref{fig:1.2mm} (a) and also than the typical widths of filaments in nearby molecular clouds, $\sim$0.1 pc (\authorcite{Doris11} \yearcite{Doris11,Doris19}).
If the width of the filament containing MC5-N/S was close to 0.1\,pc in the earlier phase as in the other filaments in nearby region, we \textcolor{black}{suggest} that radial collapse of the filament indeed occurred and resulted in the reduction of the characteristic mass of the subsequent self-gravitational fragmentation.
In this case, a blue asymmetric profile with an indication of infalling motion may be seen in optically thick lines, such as HCO$^+$ (e.g., \cite{Onishi99}). However, our HCO$^{+}$ (1--0) observation of this core with the Nobeyama 45 m telescope shows single-peaked profiles through the observed region. \textcolor{black}{Note that \citet{Seo18} also could not find collapsing motions in the HCN\,(1--0) and HCO$^{+}$\,(1--0) profile toward this core although the lack of such evidence does not necessarily mean that there are no internal inflow motions. For example, }the dynamically collapsing region is smaller than the beam size, $\lesssim$ 3,000 au, and the observed filament is not collapsing as a whole. There is a possibility that the observed line did not trace the high-density collapsing gas due to its small critical density. \textcolor{black}{Higher angular resolution observations in higher transitions of optically thick lines may provide us with the further properties of the internal motion.}\\
\ \citet{Andre12} identified a pre-brown dwarf condensation, Oph B-11, in $\rho$-Ophiuchus, which is consistent with tiny condensations predicted by the gravo-turbulent fragmentation \citep{Padoan02,Hennebelle08}. We discuss here whether turbulence played an important role in the formation and evolution of the MC5-N core.
In the large-scale view (the filamentary cloud scale), there is a possibility that the high-density condensation was formed by shock compression driven by the colliding flow.  
\citet{Tafalla15} found that two filamentary components with the relative velocity difference of $\sim$1\,km\,s$^{-1}$ are overlapping at the center of MC5-N.  Shock compressions can form the high-density portions in the postshock layer, which is more than one orders of magnitude higher than that in the preshock layer. The crossing time of the shock is calculated by dividing the size of the core ($\sim$0.03 pc) by the relative velocity difference between the two filamentary components ($\sim$1 km\,s$^{-1}$), $\sim$3 $\times$ 10$^{4}$\,yr\textcolor{black}{, which is a lower limit of the age of the core}. 
However, because dense regions formed by such low-velocity shocks can be dissipated in a short time scale \citep{Lomax16}, it is difficult to realize the high-deuterium fractionation which needs longer time scale as mentioned in Sect. 4.1. \\
\ The current 7\,m array observations also revealed an indication of substructures (i.e., multiple local peaks) in the dense core as well as the position offset between the single-dish based dust continuum peak and that obtained with the 7\,m array. The complex distribution may be explained by small-scale turbulent perturbations with a size scale of $\sim$1000 au rather than the spherically/axially symmetric collapse of the dense core. The possibility of the creation of wide multiples with the separations of more than several hundred au by turbulent perturbation within a dense core is proposed (e.g., \cite{Offner10,Tobin16}). \textcolor{black}{Although the observed linewidths of N$_2$D$^{+}$/N$_2$H$^{+}$ are indeed close to the thermal motion at the temperature of \textcolor{black}{9}\,K (Sect. \ref{R:N2HP}), the non-thermal linewidth is estimated to be $\sim$0.3\,km\,s$^{-1}$, and it is 
slightly larger than the simple extension of the standard size-linewidth relation (e.g., \cite{Solomon87}) for the length scale of 1000\,au, $\sim$0.1\,km\,s$^{-1}$. 
There is an indication that the linewidth at the center position is slightly larger than the average one within the core radius (Sect. \ref{R:N2HP}).} 
Note that complex internal substructures of dense cores originated from turbulent motions at an early phase of gravitational contraction (i.e., before protostellar core formation) tend to be unexpected in MHD simulations of Sun-like star formation (e.g., \cite{Matsumoto11}). 
In contrast, the radial collapse of a sufficiently massive filament is expected to produce many small scale ($<$0.01pc) fragments by the subsequent gravitational fragmentation but also the coalescence of the fragments is also expected depending on the actual perturbation in the filament (c.f., \cite{Inutsuka97, Masunaga99,Inutsuka01}).  Therefore this mechanism can explain both the mass of MC5-N and its possible internal substructure. \\
\ \textcolor{black}{As noted in Sect. \ref{R:1.2mm}}, because the intensity contrast between the local peaks identified by the present 7\,m array observations is small, we may be just looking at the surface \textcolor{black}{structures rather than individual gravitationally bound fragments leading to the formation of multiple stars. } 
 Higher angular resolution and high-sensitivity observations including the ALMA main array (the 12m array) are anticipated to clarify the existence of the substructures (i.e., fragments) and its formation mechanism. Early interferometric observations in molecular lines toward a starless core, TMC-1, in Taurus suggested that gravitationally unbound (possibly turbulent dominated) condensations with the size scale of 0.006 to 0.3 pc may play an important role in forming protostars by coalescence processes \citep{Takakuwa03}. Recent ALMA observations of a protostellar core, MC27/L1521F (c.f., \cite{Onishi99,Tokuda14,Tokuda17}), found high-density starless condensations with the mass of 10$^{-4}$--10$^{-3}$\,$M_{\odot}$ in the vicinity of the possible shock compressed layers originating from the internal turbulent motions within the core \citep{Tokuda18}. Future ALMA observations toward different types of dense cores will reveal to us the nature of fragmentation and collapse of very tiny condensations as precursors of tiny stellar objects. 

\section{Summary \& Future prospects}
\ We have carried out single-dish and interferometric observations with an angular resolution of $\sim$6$\arcsec$--20$\arcsec$ toward a starless core, MC5-N, located in the L1495 region in the Taurus molecular cloud. Although the total mass derived by the IRAM 30\,m telescope is very small as $\sim$\textcolor{black}{0.2--0.4}\,$M_{\odot}$, the 7\,m array observations have found a column density enhancement towards the center. In addition, the N$_{2}$D$^{+}$ and N$_{2}$H$^{+}$ observations with the Nobeyama 45 m telescope revealed a high deuterium fraction at the center of the core. We have concluded that this core is a unique source as a prestellar core on the verge of brown dwarf or very low-mass star formation. The observed physical and chemical properties may represent the initial condition of such tiny objects. \\
\ Higher angular resolution studies with the ALMA 12\,m array will be needed to investigate the actual density, fragmentation and collapse in this core. If there is higher-density gas ($\gtrsim$10$^6$\,cm$^{-3}$) at the center of the core, the present observations using the single-dishes and the 7\,m array cannot identify such higher-density condensations due to the lack of the angular resolutions.
In fact, substructures as well as a compact (a few hundred au) starless source, MMS-2, with a density of $\sim$10$^7$\,cm$^{-3}$ within a star-forming core, MC27/L1521F, in Taurus were identified for the first time using the 12\,m array (\authorcite{Tokuda14} \yearcite{Tokuda14,Tokuda16}). 
More recently, \citet{Caselli19} found a similar high-density peak in L1544 in Taurus. They indicate that such kinds of structures could be reproduced by the synthetic observations of a smoothed core; there is a possibility that the current possible substructures observed in MC5-N are also caused by the observation effect having incomplete {\it uv} coverage of the ACA observations. The follow-up 12\,m array observations will be able to cover the size scale of the substructure in the {\it uv}-plane with much better angular resolutions in order to reveal the existence and the nature of the substructure.  This follow-up study will provide us crucial hints to elucidate the nature of the innermost core.
\textcolor{black}{We note that} because the continuum flux of \textcolor{black}{high-density sources in both MC27/L1521F and L1544} is as weak as $\sim$\textcolor{black}{0.5--1}\,mJy\,beam$^{-1}$ with 1\arcsec at 1.1\,mm \citep{Tokuda14}, survey-type observations with a short integration time using the 12\,m array (e.g., \cite{Dunham16,Kirk17}) are not suitable to detect such substructures due to the low-sensitivity targeting much higher density ($\gtrsim$10$^8$\,cm$^{-3}$) objects.
\\
\ Numerical simulations by \citet{Tomida10} suggested that the lifetime of the first core, the first quasi-hydrostatic object during the star formation process (e.g., \cite{Larson69,Masunaga98,Tomida13}; \authorcite{Stamer18a} \yearcite{Stamer18a,Stamer18b}), formed in a sub-solar condensation is longer than that in typical mass (a few\,$M_{\odot}$) dense cores. MC5-N may thus be a vital object in the search for such short-lived objects. \\
\ We have unexpectedly found via our ACA survey that the sub-solar-mass condensation, MC5-N, with a lower column density compared to the other samples, is very high-density. We also emphasize that interferometric observations including short-spacing baselines are essential to further characterize the inner substructures of starless cores in nearby star-forming regions because previous single-dish studies in the dust continuum may have overlooked such structures.

\section*{Acknowledgments}

This paper makes use of the following ALMA data: ADS/ JAO.ALMA\#2016.1.00928.S. 
ALMA is a partnership of ESO (representing its member states), NSF (USA) and NINS (Japan), together with NRC (Canada), MOST and ASIAA (Taiwan), and KASI (Republic of Korea), in cooperation with the Republic of Chile. The Joint ALMA Observatory is operated by ESO, AUI/NRAO and NAOJ.
IRAM is supported by INSU/CNRS (France), MPG (Germany), and IGN (Spain). The Nobeyama 45\,m radio telescope is operated by Nobeyama Radio Observatory, a branch of National Astronomical Observatory of Japan (NAOJ). The hyper fine structure fitting of the N$_2$H$^{+}$ and N$_2$D$^{+}$ spectra (Sect. \ref{R:N2HP}) was performed with the Pyspeckit module of Python \citep{Ginsburg11}.
This work was supported by NAOJ ALMA Scientific Research Grant Numbers 2016-03B and Grants-in-Aid for Scientific Research (KAKENHI) of Japan Society for the Promotion of Science (JSPS; Grant Nos. 22244014, 23403001, 26247026, 16H05998, 18K13582, and 18H05440).

\end{document}